\documentclass[reprint,superscriptaddress,amsmath,amssymb,aps,prresearch]{revtex4-2}

\usepackage{graphicx}
\usepackage{dcolumn}
\usepackage{bm}
\usepackage{xcolor}
\usepackage{hyperref}

\begin{document}

\title{From Knots to Crystals: Machine-Learned Potentials for Self-Assembling\\ Topological Solitons in Liquid Crystals}

\author{Arunkumar Bupathy}
\email{abupathy@hiroshima-u.ac.jp}
\affiliation{International Institute for Sustainability with Knotted Chiral Meta Matter (WPI-SKCM$^2$), Hiroshima University, 1-3-1 Kagamiyama, Higashi-Hiroshima, Hiroshima 739-8531, Japan.}%

\author{Darian Hall}
\affiliation{Department of Physics, University of Colorado, Boulder, CO 80309, USA}%
\affiliation{Department of Physics, University of California, Berkeley, CA 94720, USA}%

\author{Ivan I. Smalyukh}%
\email{ivan.smalyukh@colorado.edu}
\affiliation{International Institute for Sustainability with Knotted Chiral Meta Matter (WPI-SKCM$^2$), Hiroshima University, 1-3-1 Kagamiyama, Higashi-Hiroshima, Hiroshima 739-8526, Japan.}%
\affiliation{Department of Physics, University of Colorado, Boulder, CO 80309, USA}%
\affiliation{Department of Electrical, Computer, and Energy Engineering,
Materials Science and Engineering Program, University of Colorado, Boulder, CO 80309, USA}
\affiliation{Renewable and Sustainable Energy Institute, National Renewable Energy Laboratory, University of Colorado, Boulder, CO 80309, USA}

\author{Gerardo Campos-Villalobos}
\affiliation{Soft Condensed Matter and Biophysics, Debye Institute for Nanomaterials Science, Utrecht University, Princetonplein 1, 3584 CC Utrecht, The Netherlands}%
\affiliation{CNR-ISC and Department of Physics, Sapienza University of Rome, p.le A. Moro 2, 00185 Rome, Italy}

\author{Rodolfo Subert}
\affiliation{Soft Condensed Matter and Biophysics, Debye Institute for Nanomaterials Science, Utrecht University, Princetonplein 1, 3584 CC Utrecht, The Netherlands}%

\author{Marjolein Dijkstra}
\email{m.dijkstra@uu.nl}
\affiliation{Soft Condensed Matter and Biophysics, Debye Institute for Nanomaterials Science, Utrecht University, Princetonplein 1, 3584 CC Utrecht, The Netherlands}%
\affiliation{International Institute for Sustainability with Knotted Chiral Meta Matter (WPI-SKCM$^2$), Hiroshima University, 1-3-1 Kagamiyama, Higashi-Hiroshima, Hiroshima 739-8526, Japan.}%

\date{\today}

\begin{abstract}

Knotted fields in classical and quantum systems have long been recognized for their non-trivial topologies and particle-like behavior, but practical applications have been limited by the difficulty of stabilizing them. Recently, stable knotted solitonic textures--heliknotons--were discovered in chiral liquid crystals, forming adaptive crystal assemblies via elastic distortion-mediated interactions. We use machine learning to develop single-site coarse-grained potentials that accurately capture these chiral anisotropic effective interactions. {The resulting potentials accurately reproduce experimentally observed heliknoton assemblies and enable simulations at length and time scales far beyond the range of fine-grained continuum models. This general framework is readily transferable  to other topological solitons, providing a powerful route to understand, predict, and ultimately control their collective behavior and dynamics.}

\end{abstract}

\maketitle

\section{Introduction}
In the 19th century, Gauss and Kelvin proposed that knots within fields could behave like particles, even representing atoms~\cite{Kauffman2001,Manton2004,Shnir2018}. Although this early model did not describe actual atoms, it played a pivotal role in the emergence of mathematical knot theory. Skyrme and others extended this perspective by modeling subatomic particles as topological solitons~\cite{Manton2004,Shnir2018,Skyrme1961}, while knotted field structures also emerged in classical and quantum field theories, fluid mechanics, particle physics, and cosmology~\cite{Manton2004,Shnir2018,Skyrme1961,Tai2018,Faddeev1997,Moffatt1992, Scheeler2017,Kleckner2013,Buniy2014,Shankar1977}. Singular vortex lines and topological solitons were identified as particle-like building blocks in superconductors, magnets, and liquid crystals~\cite{Chaikin1995,Han2017,Foster2019}. Yet these knotted structures were long considered as isolated objects, {often appearing as energetically unstable or metastable transient features}~\cite{Moffatt1992,Scheeler2017,Kleckner2013,Bouligand1974,Bouligand1978,Machon2014,Tai:2022}, rather than particle-like building blocks able to self-assemble into extended crystals necessary for them to behave as ordered metamatter. {Realizing such behavior requires both the stability of the topological building blocks and precise control over  the interactions between them.}

Chiral liquid crystals (CLC) host a multitude of topological structures, including skyrmions, torons, hopfions and m\"{o}biusons, stabilized by the medium's intrinsic chirality ~\cite{Smalyukh2010,Fukuda2011,Ackerman2014,Sec2014,Guo2016,Posnjak2016,Ackerman2017,Tai2019,Voinescu2020,Zhao2023}. Notably, heliknotons---Hopf solitons embedded in a helical background, or equivalently vortex knots in the immaterial helical axis field---exhibit particle-like behavior such as Brownian motion and hierarchical self-organization into crystalline structures~\cite{Tai2019,Voinescu2020}. This behavior overcomes key  barriers to forming ordered assemblies of solitonic metamatter. Unlike atomic, molecular, or colloidal crystals, the symmetry of heliknoton lattices can be tuned by small experimental adjustments~\cite{Tai2019,Voinescu2020,HallNatPhys}, making them promising platforms for metamaterial designs, information storage, unconventional computing, light modulation, and photonics. Understanding how such assemblies emerge and respond to perturbations naturally raises the question of what effective forces govern heliknoton interactions.

The forces at play manifest as strongly anisotropic interactions, with pair potentials tunable from attractive to repulsive, spanning tens to thousands of \(k_{\mathrm{B}}T\), arising from long-range perturbations of the surrounding helical fields~\cite{Tai2019}. Although these broad features are well established, the design principles of how to use these effective interactions to control resulting collective behavior remain elusive. Modeling these systems is challenging because the solitons are highly sensitive to small variations in material constants, applied fields, geometry, and boundary conditions. {This challenge is further compounded by the highly deformable nature of the underlying field configurations, particularly at higher soliton densities.} Previous studies have primarily relied on fine-grained simulations~\cite{Tai2019,Voinescu2020} and collective-variable models to describe the dynamics of individual solitonic objects~\cite{Long2021,Alvim2024,Teixeira2024a}. Particle-like descriptions are appealing as they provide a tractable framework for studying collective phenomena. Existing approaches include iterative Boltzmann inversion~\cite{Ge2023} or simplified models based on exponentials, Gaussians~\cite{Lin2013,Gonzalez2019}, or elastic multipoles~\cite{Teixeira2024}. While these methods capture the qualitative aspects of soliton dynamics and interactions, they either assume rigid textures, limiting the insights into potential reconfigurability by external stimuli, or generalize poorly to strongly anisotropic or chiral interactions.

In this work, we employ a machine learning (ML) approach to develop effective coarse-grained potentials for modeling solitonic interactions and self-assembly, using heliknotons in CLCs as a prototype. {Our results show that heliknotons behave as quasiparticles that self-organize into complex two- and three-dimensional crystal structures, closely matching experimental observations. This demonstrates the power of our approach for modeling large systems of heliknotons \textit{in-silico}, which would be prohibitively expensive using fine-grained methods. More specifically, it  provides a quantitative, particle-like description of topological solitons with deformable field configurations and strongly anisotropic interactions. Our framework is generic and can be extended to other classes of topological solitons, enabling the understanding, prediction, and control of their collective behavior and dynamics, and opening new opportunities for the design of topological metamaterials.}

\section{Results}
\begin{figure}
    \centering
    \includegraphics[width=\linewidth]{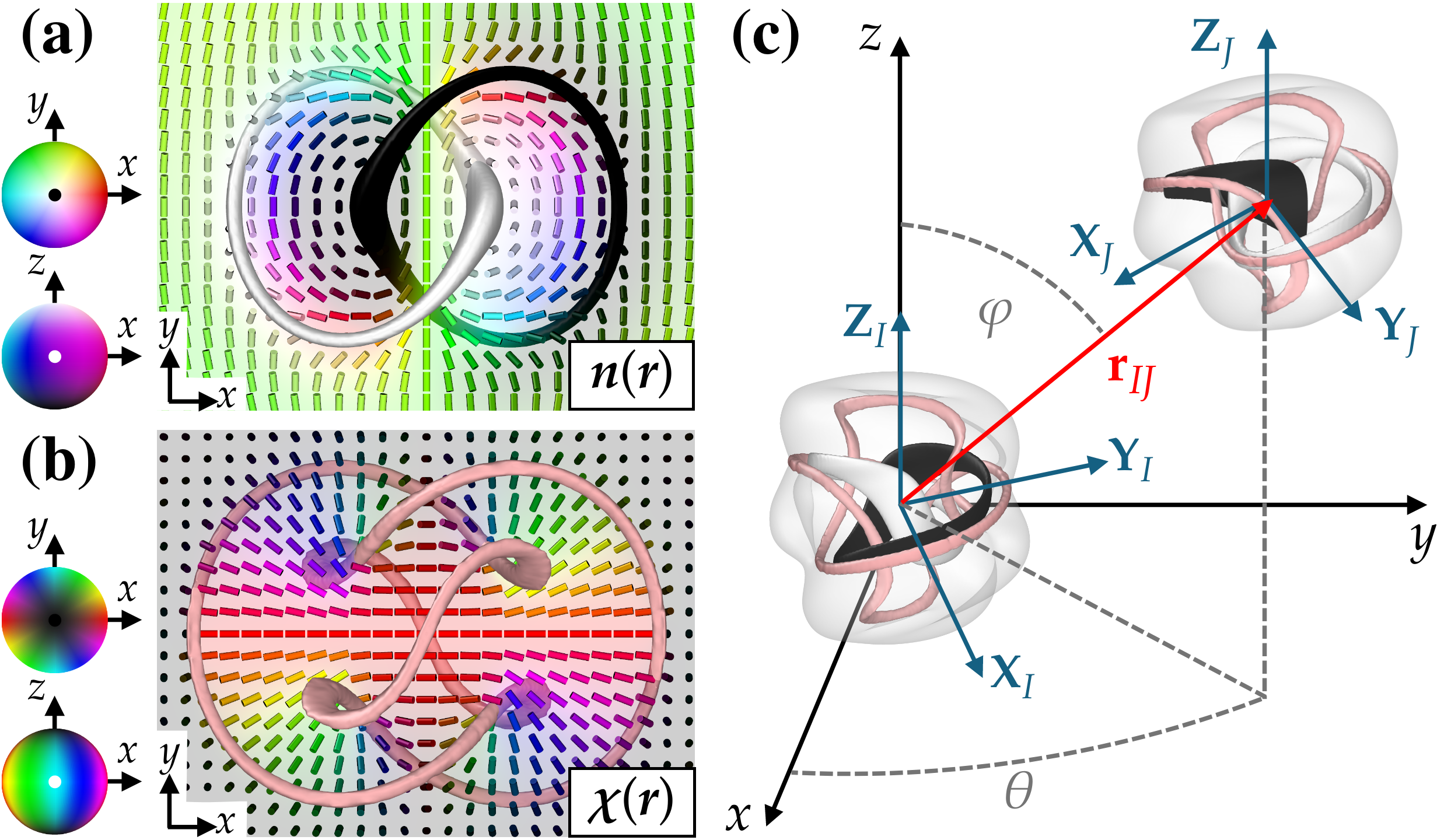}
    \caption{Cross-sections through the horizontal mid-plane showing (a) the heliknoton's director field $\boldsymbol{n}(\boldsymbol{r})$ and (b) the helical field $\boldsymbol{\chi}(\boldsymbol{r})$. The non-polar $\boldsymbol{n}(\boldsymbol{r})$ can be smoothly  vectorized without introducing singularities, as illustrated by the color scheme based on the two-sphere order parameter space of vectorized $\boldsymbol{n}(\boldsymbol{r})$. The linked loops in (a) correspond to preimages of vertical orientations in this  vectorized $\boldsymbol{n}(\boldsymbol{r})$ field. In (b), the light red tube shows singular regions in the nonpolar $\boldsymbol{\chi}(\boldsymbol{r})$ field,  forming a trefoil knot; orientations of $\boldsymbol{\chi}(\boldsymbol{r})$ are represented by a colored sphere with diametrically opposite points identified. (c) Schematic of  the position and orientation coordinates used in the coarse-grained model of heliknoton pair interactions. The gray isosurfaces surrounding the preimages and vortex knots highlight localized regions where  $\boldsymbol{\chi}(\boldsymbol{r})$ exhibits significant deviations from  its uniform far-field background.}
    \label{fig1}
\end{figure}

{\color{black}
To quantify heliknoton interactions, we first compute pair interaction energies for a range of configurations using fine-grained (FG) simulations based on the Frank–Oseen free-energy functional, which captures the elastic cost of distortions in the nematic director field within a confined cholesteric medium. These energies are then used to train a coarse-grained (CG) model in which each heliknoton is represented as an effective particle defined by its geometric center and orientation, as illustrated in Fig.~\ref{fig1}. The interaction potential is expressed as a linear expansion in symmetry functions that are descriptors of the local structure, following Ref.~\cite{Campos-Villalobos2024}, with parameters optimized to reproduce the FG energies. In particular, the orientational dependence is represented using \textit{S-functions} which form a complete basis set capable of capturing arbitrary angular variations~\cite{Blum1972,Stone1978}. This representation is fully differentiable and thus provides direct access to forces and torques, thereby enabling efficient dynamical simulations. A concise description of the construction is provided in the Appendix~\ref{theappendix}, while a detailed account can be found in Ref.~\cite{Campos-Villalobos2024}. The FG simulations were performed using elastic constants chosen to match those of the experimental liquid crystal mixtures LC-1 and LC-2 reported in Refs.~\cite{Tai2019,Tai2020} (see Table~\ref{tab1}, Appendix~\ref{theappendix}).}

We validate our ML model on thin LC cells, where heliknotons are effectively localized near the horizontal mid-plane for cell thicknesses $d<3p$ and large electric fields~\cite{Tai2019}. In this case, the heliknoton pair interactions are only a function of their in-plane separation $\Delta x, \Delta y$, as they share a common orientation which is a function of their $z$-coordinates (see Appendix~\ref{theappendix}). Fig.~\ref{fig2}(a) shows the two-dimensional pair interaction potential of heliknotons in the LC-1 mixture, for cell thickness $d=3p$ and applied voltage $U = 2.2$V. At separations $r=\sqrt{\Delta x^2+\Delta y^2} > 5p$, the interactions decay to zero. {Even in the simplest case of 2D confinement,  the pair interactions between  heliknotons are strongly anisotropic.} Fig.~\ref{fig2}(b) shows the radial profiles of the interaction potential along different $\theta = \tan^{-1}(\Delta y/ \Delta x)$. The CG model $\Phi$ (lines) accurately reproduces the anisotropic FG interaction potential $\Delta F$ (symbols), with a root-mean-squared error (RMSE) of $3.9k_{\mathrm{B}}T$ ($T=293$K), much smaller than the  total interaction-strength range ($\sim 6000 k_{\mathrm{B}}T$).
\begin{figure*}
    \centering
    \includegraphics[width=0.8\linewidth]{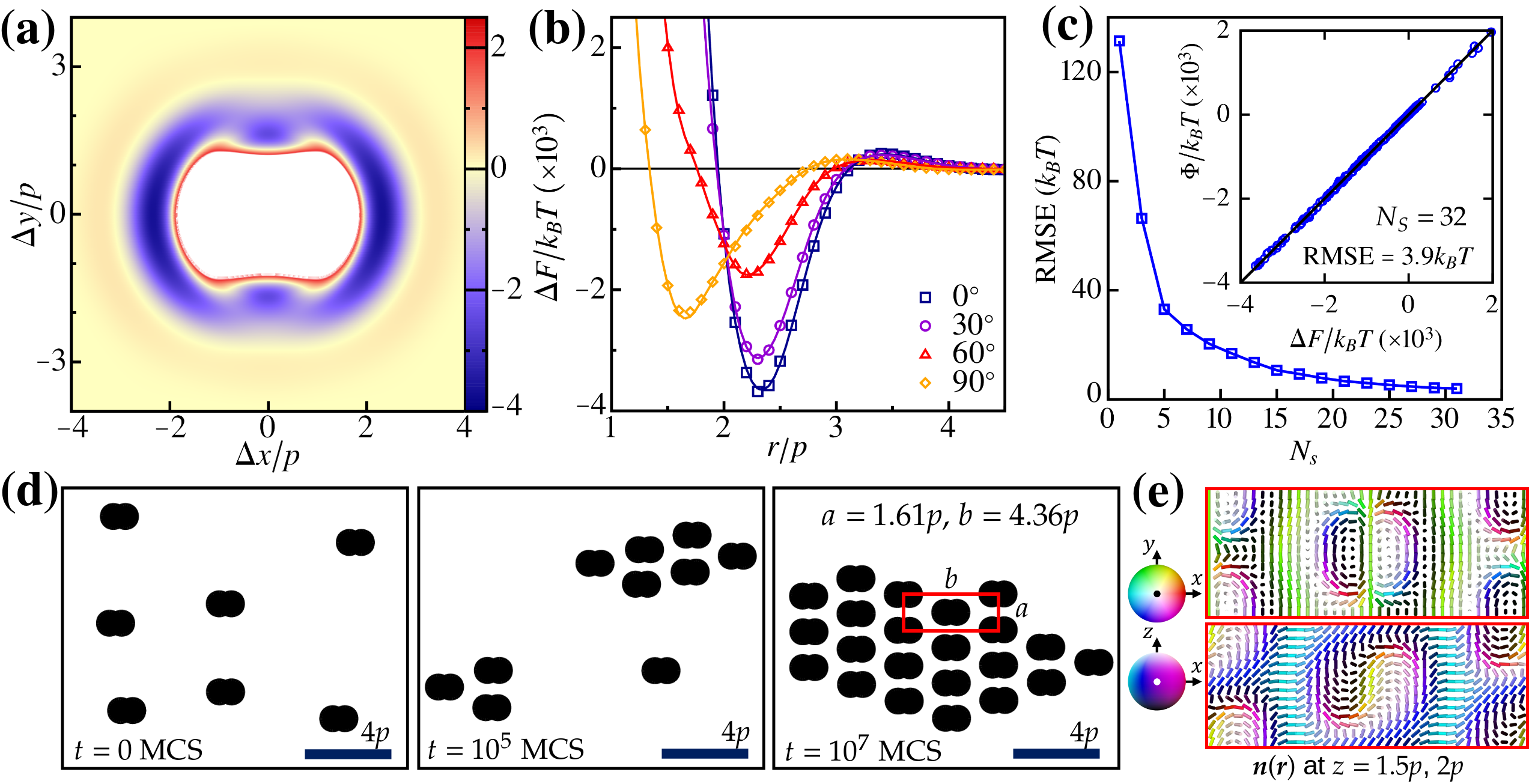}
    \caption{Results for LC-1, $d=3p$, and $U=2.2$V: (a) Pair interaction potential of heliknotons confined to the horizontal mid-plane of the LC cell, obtained from fine-grained (FG) simulations, shown as a color map. (b) Radial profiles of the FG pair potential $\Delta F$ (symbols) compared with the coarse-grained (CG) model $\Phi$ (lines) along different angles $\theta = \tan^{-1}(\Delta y/ \Delta x)$. {(c) Convergence of training errors as a function of  the number of terms $N_s$ in the CG model. The inset shows a parity plot comparing the sampled potential $\Delta F$ with the trained CG model $\Phi$.} (d) Heliknoton assemblies at successive stages of a Monte Carlo (MC) simulation using the CG model; the black symbols represent heliknoton cores and time is measured in MC sweeps (MCS). (e) Horizontal cross-sections of $\boldsymbol{n}(\boldsymbol{r})$ at different heights within a unit cell from FG simulations; the color spheres indicate the local orientations of $\boldsymbol{n}(\boldsymbol{r})$.}
    \label{fig2}
\end{figure*}

{Fig.~\ref{fig2}(c) shows the convergence of the training error as a function of the number of terms $N_s$ used in the CG model. As shown by the parity plot in the inset ($N_s = 32$), we obtain excellent agreement between the model predictions ($\Phi$) and the pair potential ($\Delta F$) sampled from FG simulations.} Using the trained CG potential $\Phi$, we then simulate the self-assembly of heliknotons via Monte Carlo (MC) simulations~\cite{Whitelam2007,Whitelam2009,VMMC} (see Appendix~\ref{theappendix}). Fig.~\ref{fig2}(d) shows snapshots of the heliknoton assemblies at successive stages of the simulation. The resulting structure that self-assembled in our simulations is a rhombic crystal with a centered rectangular unit cell, in close agreement with  the experimental observations~\cite{Tai2019}. Although the apparent self-assembled lattice of particle-like solitons is rectangular, the crystal retains the chirality of the underlying chiral nematic host medium and heliknoton textures, as evident from Fig.~\ref{fig2}(e), which show horizontal sections of $\boldsymbol{n}(\boldsymbol{r})$ obtained from FG simulations of a unit cell.

\begin{figure*}
    \centering
    \includegraphics[width=0.8\linewidth]{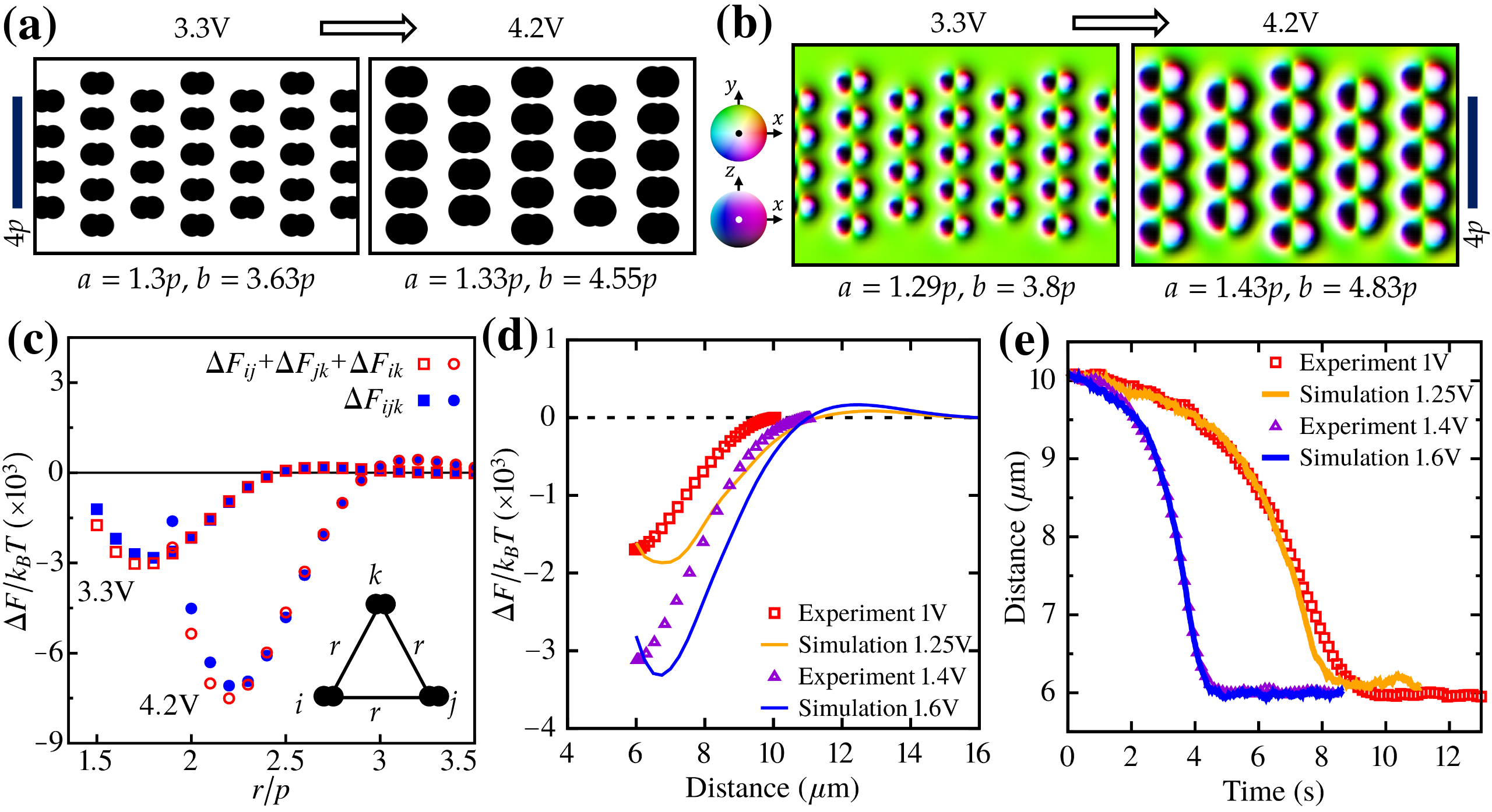}
    \caption{(a--c) Results for LC-2, $d=2p$: (a) Electrostriction of a heliknoton crystal obtained from MC simulations, following a voltage change from $U = 3.3$V to $4.2$V. (b) Corresponding horizontal sections of the director field obtained from FG simulations following the voltage change. (c) Comparison of three-body energies versus the sum of pairwise energies from FG simulations. {(d--e) Results for LC-1, $d=2p$: (d) Heliknoton pair potentials from experiments (symbols) compared to the one-dimensional section of the CG potentials obtained from simulations (lines) (e) Time evolution of heliknoton pair separation obtained in experiments compared to simulations. The simulation data has been scaled to fit the experiments. The experimental data are from Ref.~\cite{Tai2019}.}}
    \label{fig3}
\end{figure*}

Subsequently, we test our ML approach to capture small changes in experimental conditions by training the models on data from FG simulations in the LC-2 mixture ($d=2p$) at electric field strengths of $3.3$V and $4.2$V. We first perform MC simulations using the CG potential at $3.3$V, to identify the crystal lattice self-assembled by the heliknotons. Initializing a crystallite in this state (Fig.~\ref{fig3}(a), left), we then switch the interaction potential to that at $4.2$V and observe its expansion (Fig.~\ref{fig3}(a), right). FG simulations using the Frank–Oseen functional (Fig.~\ref{fig3}(b)) yield slightly different lattice constants, likely reflecting the absence of many-body interactions in the CG model. To test this, we compare the three-body energy $\Delta F_{ijk}$ of a heliknoton triplet $(i,j,k)$ arranged in an equilateral triangle of length $r$ versus the sum of pairwise energies $\Delta F_{ij}$, $\Delta F_{jk}$, $\Delta F_{ik}$ in Fig~\ref{fig3}(c). For separations $r > 2.5p$, we observe good agreement between the full three-body potential and the sum of two-body terms. At smaller separations, although the pairwise sum underestimates the total interaction energy, the qualitative behavior is still successfully captured. Nonetheless, the CG simulations are highly efficient, completing in about $2.7$ hours on four Intel i9-13900 CPU cores, compared to $28.1$ hours for the FG simulations on an Nvidia RTX A4000 GPU.

{Because the pair potentials obtained from our ML model depend on both heliknoton separation and orientation, we can straightforwardly compute the resulting forces and torques acting on the heliknotons and  simulate their dynamics. We demonstrate this for a pair of heliknotons initially separated by $2.4p$  along the $y$-axis, a configuration where both  forces and torques  are close to zero. Fig.~\ref{fig3}(d) shows a one-dimensional section of the machine-learned pair potential along the $y$-axis for two different voltages $U=1.25$V and $1.6$V, at a cell thickness $d=2p$ in LC-1 (solid lines), together with the corresponding experimentally obtained pair potentials at $U=1$~V and $1.4$V (symbols) from Ref.~\cite{Tai2019}. Fig.~\ref{fig3}(e) shows the time evolution of the separation distance of the heliknotons obtained from Brownian dynamics simulations (see Appendix~\ref{theappendix}) using the ML potentials, rescaled to match the  experimental data, as discussed below.}

{It may be noted that although the well depths are roughly matched, the simulation voltages are about $0.25$V higher than in the experiments, and the positions of the minima and maxima are shifted by about $10\%$. This small discrepancy is, however, not a limitation of the ML method, as the RMS errors of the ML potentials are only $\sim 0.2\%$ of the total pair-interaction range sampled from the FG simulations. Rather, it arises from simplifications employed in the Frank-Oseen calculations used to sample the interaction potential, in particular the omission of the electrostatic nonlocal field effects (related to rather large dielectric anisotropy of the liquid crystal medium) and the saddle-splay ($K_{24}$) contributions. Although there is room to improve the fine-grained simulations, this lies beyond the scope of the present work. Nevertheless, when the dynamical trajectories from simulations are rescaled to account for the shifts in the potential mentioned above, we find good agreement with the experimentally observed structures and dynamics.}

\begin{figure*}
    \centering
    \includegraphics[width=0.9\linewidth]{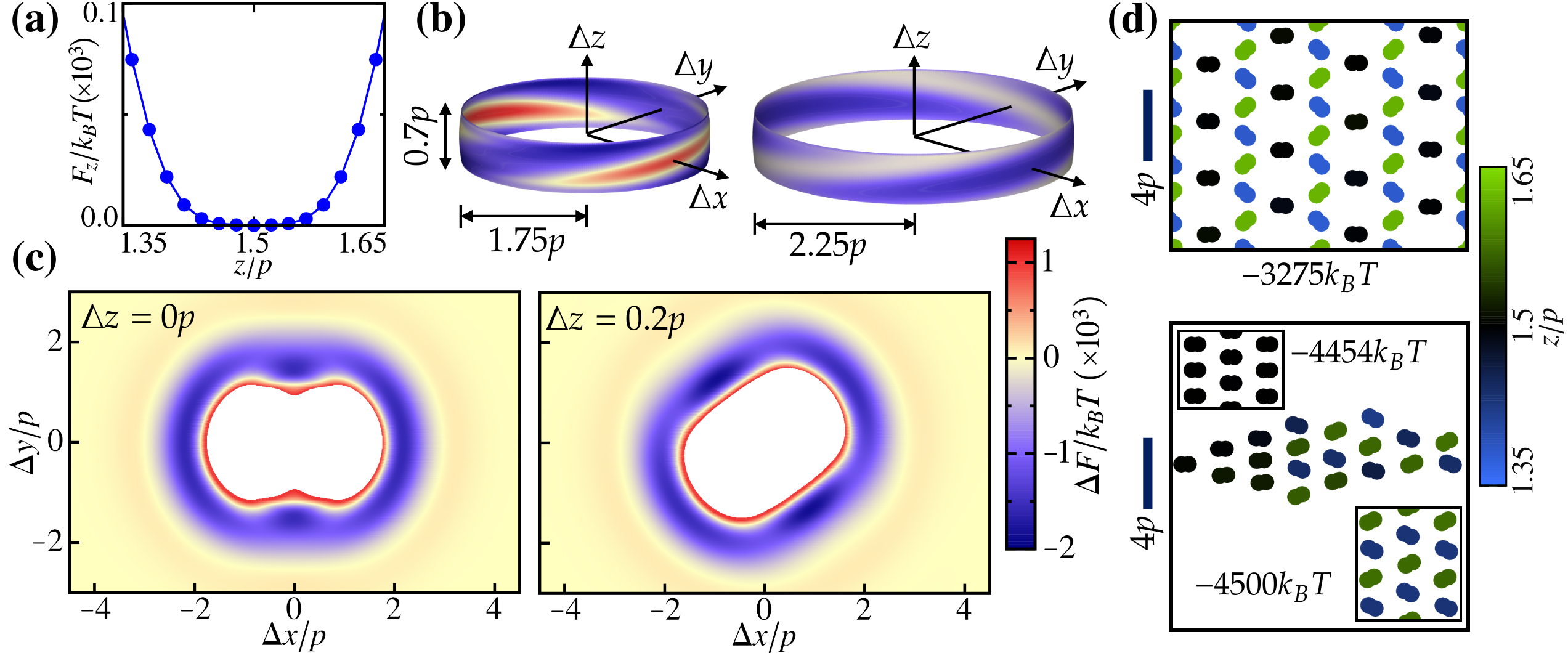}
    \caption{Results for LC-1, $d=3p$ and $U=2.08$V: (a) $z$-dependence of the potential experienced by an individual heliknoton. (b) Three-dimensional heliknoton pair potential shown for separation distances $1.75p$ and $2.25p$, shown as color maps on spherical shells truncated by the sampled vertical range. (c) Horizontal sections of the pair potential at different vertical separations $\Delta z$. (d) Snapshots from MC simulations showing the stability of a stretched kagome crystal (top) and self-assembly into rhombic crystals (bottom); colors represent the $z$-position of the heliknotons. The insets show the two observed rhombic crystals (not unit cells) along with the corresponding energy per heliknoton.}
    \label{fig4}
\end{figure*}

{\color{black}{We next consider the experimentally realized case in which heliknotons} form a stretched kagome lattice with staggered $z$-positions (LC-1, $d=3p$, $U=1.7$V)~\cite{Tai2019}. In our simulations, this configuration is obtained at $U=2.08$V.} In this setting, it is also necessary to sample the energy of a single heliknoton as a function of its $z$-position, since individual heliknotons interact with the LC cell surfaces. The $z$-potential shown in Fig.~\ref{fig4}(a) exhibits a wide minimum around the vertical midsection ($1.5p$) of the cell. Fig.~\ref{fig4}(b) shows the three-dimensional heliknoton pair potential for separation distances $1.75p$ and $2.25p$, as color maps on spherical shells truncated by the sampled vertical range, whereas  Fig.~\ref{fig4}(c) presents the corresponding horizontal sections of the potential at different vertical separations $\Delta z$. The twist observed in Figs.~\ref{fig4}(b,c) clearly shows that the interactions are not only anisotropic but also chiral.

MC simulations using the CG model confirm the stability of the stretched kagome crystal with staggered $z$-positions (top panel, Fig.~\ref{fig4}(d)), as observed experimentally~\cite{Tai2019}, when initialized in this configuration. The opposite-handed configuration, with the $z$-positions swapped, is higher in energy by about $330k_{\mathrm{B}}T$ and unstable. Self-assembly from a dilute gas, however, produces rhombic crystals, with the heliknotons either near the horizontal mid-plane or adopting staggered z-positions (bottom panel, Fig.~\ref{fig4}(d)), the latter being lowest in energy. {The discrepancy with experiments likely arises from simplifications  in the FG simulations, as already discussed in the context of Fig.~\ref{fig3}. The ML-derived potential  reproduces the pair-interaction landscape sampled using the FG simulations with an error of only $\sim 0.2\%$.}

{Our method is not limited to $2D$ systems and is sufficiently general, capable of simulating $3D$ crystals. Fig.~\ref{fig5}(a) shows the $z$-dependence of the potential experienced by a single heliknoton at $U=3.8$V for a cell thickness of $6p$ in LC-1. The $z$-potential exhibits sinusoidal oscillations with a period of $0.25p$ with large barriers appearing as the heliknoton approaches the top or bottom surfaces of the cell. We fit this potential using a combination of sinusoidal terms and even polynomials in $z$.
\begin{figure*}
    \centering
    \includegraphics[width=0.8\linewidth]{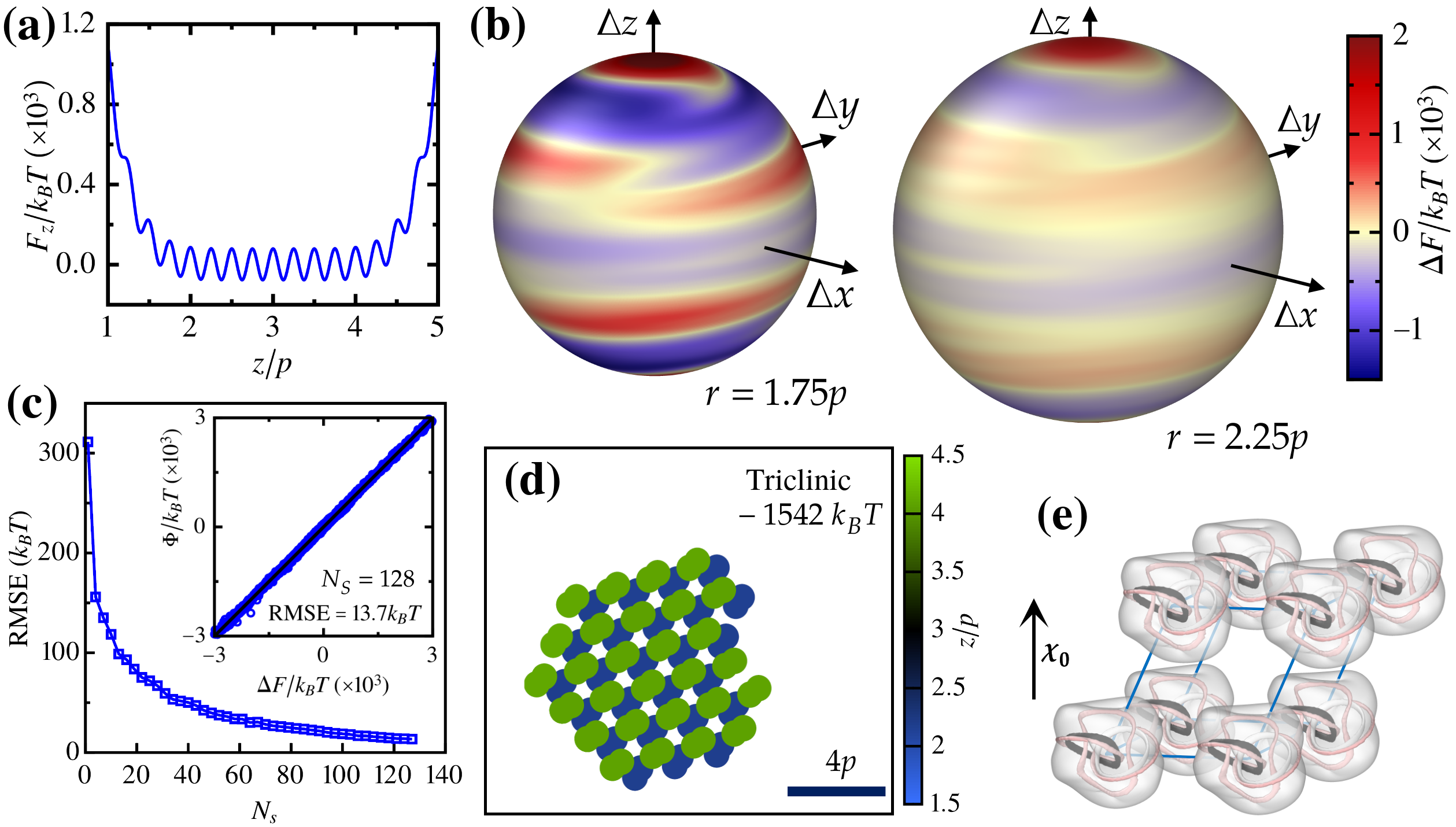}
    \caption{Results for LC-1, $d=6p$, $U=3.8$V: (a) $z$-dependence of the potential experienced by an individual heliknoton. (b) Three-dimensional heliknoton pair potential shown for separation distances $1.75p$ and $2.25p$, shown as color maps on spherical shells. (c) Convergence of training errors with the number of terms $N_s$ in the CG model. The inset shows the parity plot comparing the sampled potential $\Delta F$ with the trained CG model $\Phi$. (d) Snapshots from MC simulations showing the 3D assembly obtained using the ML potential $\Phi$ along with the corresponding energy per heliknoton; colors represent the $z$-position of the heliknotons. (e) Visualization of the unit cell of the triclinic crystal obtained from simulations using the ML potential. $\boldsymbol{\chi}_0$ is the far-field helical axis.}
    \label{fig5}
\end{figure*}

In Fig~\ref{fig5}(b), we show the corresponding $3D$ interaction potential of heliknoton pairs at two separation distances $r=1.75p$ and $2.25p$, with the interaction strength represented by a color map. This corresponds to  a total $z$-displacement range of $3.5p$ and $4.5p$, respectively, giving rise to pronounced chiral features in the pair interactions that depend on the $z$-separation of the heliknotons. Although more challenging than the earlier cases, the ML model nevertheless converges with the inclusion of additional  terms, reaching an RMS error of $\sim 13.7k_{\mathrm{B}}T$ for a model with $N_s = 128$ terms, as shown in Fig.~\ref{fig5}(c). This corresponds to an error of less than $0.25\%$ of the total interaction-strength range.}

{MC simulations using the trained CG potential $\Phi$ first lead to the self-assembly of monolayers of heliknotons into rhombic crystals, which, when guided together as done in experiments~\cite{Tai2019}, form triclinic crystals. In the triclinic crystal, the heliknotons are co-aligned, as seen in Fig.~\ref{fig5}(e), with a vertical separation of $\sim 2p$, qualitatively matching the experimental observations~\cite{Tai2019}, despite the simplifications used to generate the training data.}
Interestingly, inspection reveals that the model contains the $S$-function (dot-product) pairs $(\boldsymbol{X}_{I}\cdot\hat{\boldsymbol{r}}_{IJ})(\boldsymbol{X}_{J}\cdot\hat{\boldsymbol{r}}_{IJ})$ and $(\boldsymbol{Y}_{I}\cdot\hat{\boldsymbol{r}}_{IJ})(\boldsymbol{Y}_{J}\cdot\hat{\boldsymbol{r}}_{IJ})$, which, although achiral by themselves, inherit chirality due to the coupling of $\boldsymbol{X}_{I,J}$ and $\boldsymbol{Y}_{I,J}$ with the helical background (see Fig.~\ref{figs1}, Appendix~\ref{theappendix}), via a sinusoidal (odd) contribution in the vertical separation of heliknotons.

\begin{figure*}
    \centering
    \includegraphics[width=0.7\linewidth]{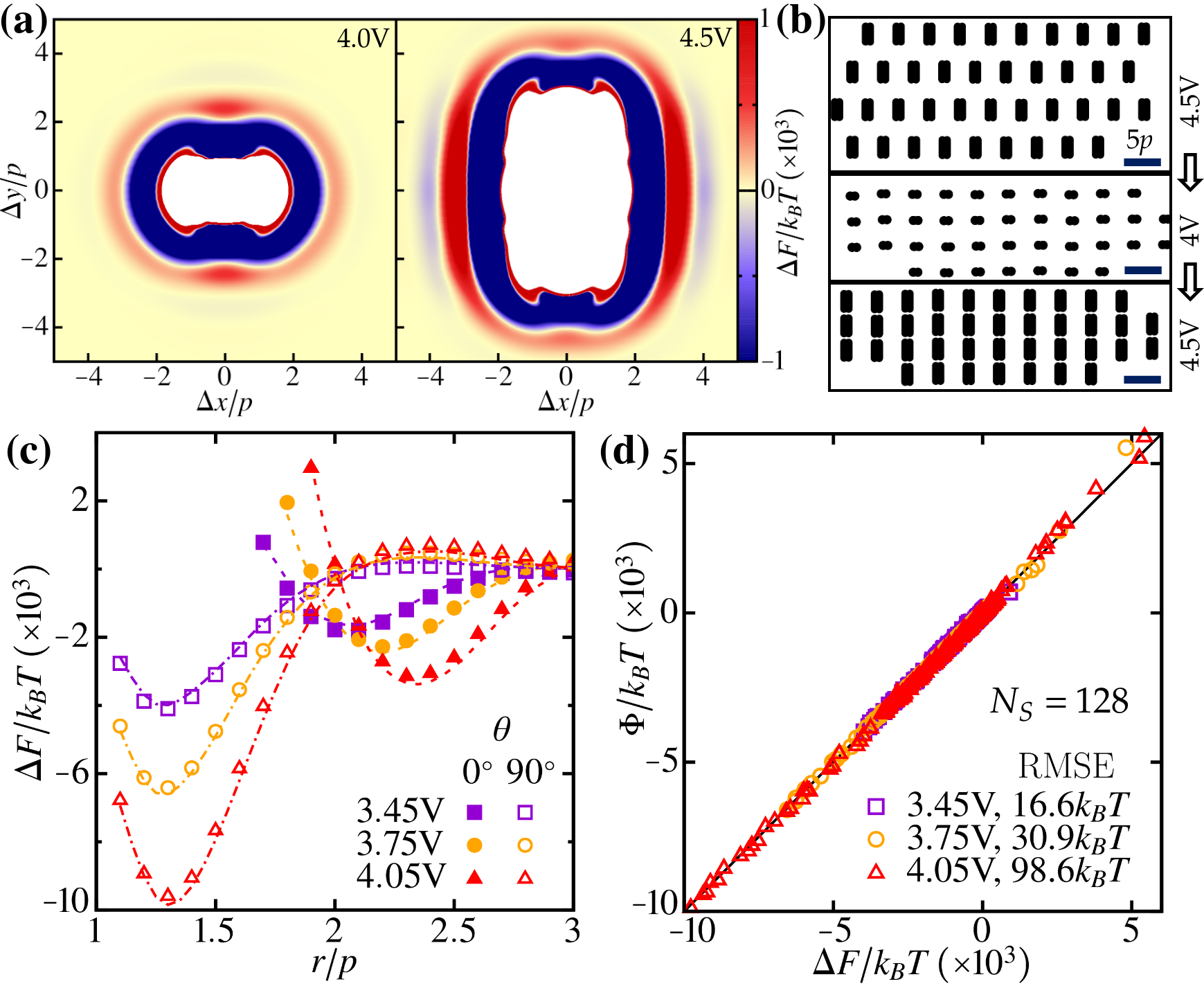}
    \caption{Results for LC-2, $d=2p$: (a) Two-dimensional heliknoton pair potentials at voltages $U=4$V and $4.5$V shown as color maps. (b) Heliknoton assemblies from MC simulations using the CG models, as the voltage is cycled from $4.5$V to $4$V and back. (c) Interpolated heliknoton pair potentials (dotted dashed lines) compared with those from FG simulations (symbols) at different applied voltages. (d) Parity plots comparing the interpolated potentials ($\Phi$) versus the ground truth from the FG simulations ($\Delta F$) along with the RMS errors.}
    \label{fig6}
\end{figure*}

To demonstrate how the ML-based CG model improves search efficiency, we present an example where a simple cycling of the applied voltage reconfigures a  heliknoton assembly. Fig.~\ref{fig6}(a) shows the two-dimensional pair potential of heliknotons in LC-2, for cell thickness $d=2p$ at applied voltages $U=4$V and $4.5$V. We observe a substantial change in the anisotropy of the interactions, with increased energy barriers. When allowed to self-assemble at $4.5$V the heliknotons form a rhombic crystal (top panel, Fig.~\ref{fig6}(b)). Cycling the voltage from $4.5$V  down to $4$V and back drives the heliknotons into a stable rectangular lattice (middle and bottom panels of Fig.~\ref{fig6}(b)), {highlighting how the ML-based CG model enables rapid exploration of assembly pathways that would be computationally prohibitive with fine-grained simulations.}

Crucially, our method is not limited to the discrete voltages at which the CG models are trained, but enables interpolation across voltage values. Given the CG potentials trained at a set of voltages $\{U_i\}$, expressed as $\Phi_i = \sum_{k_i} w_{k_i}^0 G_{k_i}$, where $w_{k_i}^0$ and $G_{k_i}$ are the coefficients and symmetry functions at $U_i$, we  construct a voltage dependent model: $\Phi(U) = \sum_i\sum_{k_i} w_{k_i}(U) G_{k_i}$. The voltage-dependent weights $w_{k_i}(U)$ satisfy $w_{k_i}(U_i) = w_{k_i}^0$ and can be interpolated using standards methods, provided the training voltages $U_i$ are sufficiently close. As an example, we use  cubic spline interpolations of  the weights $w_{k_i}(U)$ to estimate  interaction potentials at $3.45$V, $3.75$V and $4.05$V from CG models trained at $3.3$V, $3.6$V, $3.9$V and $4.2$V, each containing $N_s=32$ terms, giving a total of $128$ terms in the interpolated model, for LC-2 cells with thickness $d = 2p$. As shown in Fig.~\ref{fig6}(c), the interpolated potentials (dash-dotted lines) accurately reproduce the radial and angular dependence of the interactions observed in FG simulations (symbols) across the entire interpolation range. Fig.~\ref{fig6}(d) compares the interpolated potentials ($\Phi$) against the ground truth ($\Delta F$) obtained from  FG simulations. The RMS errors range from about $17\,k_{\mathrm{B}}T$ to $100\,k_{\mathrm{B}}T$, which are relatively small compared to the interaction scale of  $\sim 10^4\,k_{\mathrm{B}}T$. This framework can be further extended to interpolate over material constants, enabling efficient exploration of a broader parameter space without retraining the CG potentials.

\section{Discussion}

In summary, our work establishes a general framework for modeling topological solitons as quasiparticles described solely by their geometric centers and orientations. By constructing machine-learned coarse-grained potentials, we demonstrate the self-assembly of heliknotons into complex closed and open crystal lattices, in qualitative agreement with experiments. {\color{black}This approach enables dynamical simulations over length and time scales far beyond the reach of fine-grained continuum methods. Importantly, because the model uses particle-centered descriptors of local structure as regressors in a linear framework, the resulting interaction potentials remain physically interpretable, in contrast to typical black-box machine-learning approaches.}
{\color{black}Remarkably, this representation and approach remain accurate even when the heliknoton textures are highly deformed, as in the dense assemblies considered here. Although this regime involves many-body interactions, our results show that a pair-interaction-level representation can nevertheless capture the essential physics despite the complex underlying field configurations.}

{\color{black}This framework can also be extended to interpolate interaction potentials across system parameters, enabling efficient exploration of large material and control-parameter spaces, as demonstrated in our voltage-dependent analysis. Such capabilities open the possibility of systematically discovering new collective phases and reconfiguration pathways in assemblies of topological solitons.} % A natural next step will be to extend this approach to nonequilibrium situations where interactions themselves evolve in response to external driving, as explored for related systems by Teixeira et al.~\cite{Teixeira2024}.
{\color{black}More generally, the presented framework is readily transferable to other topological textures across a wide range of nonlinear and topological systems, opening new opportunities to investigate collective behavior, phase transitions, and programmable self-assembly in systems ranging from soft condensed matter to magnetic and quantum materials that host particle-like topological excitations~\cite{Liu2018,Shnir2021,Baizakov2015,Yu2010,Leonov2026,Zhao2023b,Voinescu2020}.}

{The use of our computational approach may aid in the design of light-matter interactions and photonic crystals and optical metamaterials based on topological solitons ~\cite{PoyNatPhoton, CaoNatMater}. Since the crystalline and other self-assemblies of particle-like topological solitons can be accurately predicted for a vast parameters space corresponding to different materials, confinement and surface boundary conditions and applied fields, these predictions can be used to enable steering and localization of laser light by topological solitonic meta-atoms, which may find technological utility in wide-angle beam-steering, optical energy storage and electro-optic applications. While chiral nematic and blue phases are known to be capable of acting as, respectively, one-dimensional and three-dimensional photonic crystals ~\cite{CaoNatMater}, the computationally guided self-assembly of various crystalline lattices of heliknotons and other topological solitons may allow for new forms of self-assembled and highly reconfigurable photonic metamaterials, including for the visible spectral range when lattice parameters of solitonic crystals can be brought to the submicrometer range.

From a more fundamental perspective, the facile learning of how topological solitons and knotted vortices self-organize into crystalline arrays may provide computational and experimental insights into vortex knot reconnections and topological invariant transformations during various knot fusion and fission processes that can occur within crystalline arrays ~\cite{HallNatPhys, Kleckner2013}, potentially providing pure and applied math inspirations.}

\section*{Acknowledgments}
A.B. acknowledges the N8 Centre of Excellence in Computationally Intensive Research (N8 CIR) funded by the N8 research partnership and EPSRC (Grant No. EP/T022167/1), coordinated by the Universities of Durham, Manchester and York, for access to computational resources. I.I.S. and D.H. acknowledge the support by the U.S. Department of Energy, Office of Basic Energy Sciences, Division of Materials Sciences and Engineering, under contract DE-SC0019293 with the University of Colorado at Boulder. A.B., I.I.S. and MD. acknowledge the International Institute for Sustainability with Knotted Chiral Meta Matter (WPI-SKCM$^2$) under Japan's World Premier International Research Center Initiative that made this collaboration possible.

\section*{Author contributions:}
I.I.S. and M.D. initiated and supervised the research project. I.I.S., M.D., D.H., R.S., and G.C.-V. conceptualized the project. All authors contributed to the development of the methodology. A.B., D.H. and I.I.S. developed and refined the approach and codes for fine-grained pair interaction modeling based on the Frank-Oseen free energy functional. A.B. performed the simulations, analyzed and visualized the data with input from all authors. A.B. and M.D. co-wrote the original draft. All authors contributed to the review and editing of the manuscript. Funding acquisition: I.I.S. and M.D.

\section*{Competing interests:}
There are no competing interests to declare.

\section*{Data Availability:}
All data supporting the findings can be found at~\cite{BupathyData2026}. The simulation codes are available from the corresponding authors upon reasonable request.

\appendix
\renewcommand{\thesection}{}
\section{Methods}
\label{theappendix}

\subsection{Frank-Oseen free energy calculations}

The interaction energies of heliknoton pairs are obtained from fine-grained simulations based on the Frank-Oseen free-energy functional, which describes the energetic cost of spatial deformations of the nematic director field $\boldsymbol{n}(\boldsymbol{r})$:
\begin{multline}
    F_{\mathrm{bulk}} = \! \int \! \mathrm{d^3}r \left\{ \frac{K_{11}}{2}(\nabla \cdot \boldsymbol{n})^2 + \frac{K_{22}}{2}\left[\boldsymbol{n}\cdot(\nabla\times\boldsymbol{n}) +\frac{2\pi}{p}\right]^2 \right. \\
        \left. + \frac{K_{33}}{2}\left[\boldsymbol{n}\times(\nabla\times\boldsymbol{n})\right]^2 - \frac{\epsilon_{0}\Delta\epsilon}{2}(\boldsymbol{E}\cdot\boldsymbol{n})^2 \right\},
\end{multline}
where $K_{11}$, $K_{22}$ and $K_{33}$ are the Frank elastic moduli associated with splay, twist, and bend deformations, respectively, $p$ is the cholesteric pitch, $\epsilon_0$ is the vacuum permittivity, and $\Delta\epsilon$ is the dielectric anisotropy. {We neglect the contribution of saddle-splay ($K_{24}$) deformations, as no singular defects are present in our system, the surface boundary conditions exhibit strong anchoring, and prior work has shown that including the $K_{24}$ term does not significantly affect the  results in such cases~\cite{Tai2019, PoyNatPhoton}}.

The surface anchoring energy follows the Rapini-Papoular form~\cite{Rapini1969}:
\begin{equation}
    F_{\mathrm{surface}} = - \int \mathrm{d^2}r \frac{W}{2} (\boldsymbol{n} \cdot \boldsymbol{n}_0)^2,
\end{equation}
where  $W$ is the anchoring strength and $\boldsymbol{n}_0 = \boldsymbol{y}$ is the easy axis, corresponding to unidirectional planar anchoring. We employ periodic boundary conditions along the $x$ and $y$ directions. 

The free-energy functional is numerically minimized using a variational relaxation routine employing second-order finite differences. At each step $\boldsymbol{n}(\boldsymbol{r})$ is updated as $n_i^{\mathrm{new}} = n_i^{\mathrm{old}} - \alpha \Delta t\left[F\right]_{n_i}$, where $\left[F\right]_{n_i}$ is the functional derivative of $F$ with respect to the component $n_i$~\cite{Tai2020}, and $\mathrm{\Delta t} = \Delta h^2/[2\max(K)]$ is the maximum stable time step, where $\Delta h$ is the computational grid spacing and $\max(K)$ is the largest elastic constant~\cite{Ackerman2017}, and $0\leq \alpha \leq 1$ is a reduced relaxation rate. $\boldsymbol{n}(\boldsymbol{r})$ is normalized after each update.

We initialize a pair of heliknotons $(I,J)$ at a prescribed separation vector $\boldsymbol{r}_{IJ} = (\Delta x, \Delta y, \Delta z)$, followed by numerical minimization of the free energy. The pair interaction potential is then computed as
\begin{equation}
\Delta F = F_{2} - 2 F_{1} + F_{0},
\end{equation}
where $F_{2}$ is the free energy (FE) of the relaxed dimer configuration, $F_{1}$ is the free energy of a single isolated heliknoton, and $F_{0}$ is the free energy of the uniform cholesteric state. Simulations were performed for the liquid crystal mixtures LC-1 and LC-2 studied in Ref.~\cite{Tai2019}, with the material parameters listed in Table.~\ref{tab1}, and with helical pitch $p = 5\mu$m and surface anchoring strength $W = 10^{-4}$~J/m$^2$.
\begin{table}
    \centering
    \begin{tabular}{|c|c|c|c|c|}
    \hline
        Mixture & $K_{11}$ (pN) & $K_{22}$ (pN) & $K_{33}$ (pN) & $\Delta\epsilon$ \\
    \hline
        LC-1 & 6.4 & 3 & 10 & 13.8\\
        LC-2 & 14.1 & 6.7 & 15.5 & 3.4 \\
    \hline
    \end{tabular}
    \caption{Material parameters of the LC mixtures used in the Frank-Oseen calculations.}
    \label{tab1}
\end{table}

In our simulations, both the external field $\boldsymbol{E}$ and the far-field helical axis $\boldsymbol{\chi}_0$ were aligned along the $z$-direction and non-local field effects were excluded. The simulation volume was discretized into a computational grid with $32$ points per pitch $p$, per axis. The $x$ and $y$ dimensions of the simulation cell were taken sufficiently large ($L \geq 12p$) to minimize interactions with the periodic images. For a given set of parameters, a single heliknoton is first generated using an ansatz~\cite{HallNatPhys,Kuchkin_2023} and relaxed for $64000$ steps at a relaxation rate $\alpha = 0.35$. This \textit{pre-relaxed} ansatz is then used to initialize the pair configurations which were subsequently relaxed for a further $2000$ steps to obtain the pair potential. For LC-2 at $4.5$V, the heliknoton ansatz was pre-relaxed for $64000$ steps with a surface anchoring strength $W = 10^{-6}$~J/m$^2$ followed by another $16000$ steps at $W=10^{-4}$~J/m$^2$, prior to sampling the interaction potential. {The pair interactions were sampled on a cylindrical grid, with a radial resolution of $0.1p$, azimuthal resolution of $7.5^{\circ}$, and vertical resolution of $0.05p$. This corresponds to approximately $2400$ data points for the 2D cases and about $200000$ points for the 3D case shown in Fig.~\ref{fig5}.}

\subsection{Machine-Learned Coarse-Grained Potentials}

Our CG model treats each heliknoton as an effective particle, with the total energy of the system given by  the sum of pairwise interactions. A dimer configuration is characterized by the center-to-center separation $r_{IJ} = |\boldsymbol{r}_{IJ}|$, and the angular variables $\Omega_i \equiv \Omega_i(\boldsymbol{X}_i,\boldsymbol{Y}_i,\boldsymbol{Z}_i)$, where $\boldsymbol{X}_i$, $\boldsymbol{Y}_i$ and $\boldsymbol{Z}_i$ are the body-fixed axes representing the orientation of the heliknoton $i \in \{I,J\}$, as well as $\Omega_{IJ} = (\varphi,\theta)$, describing the orientation of $\boldsymbol{r}_{IJ}$. For heliknotons embedded in a CLC with planar anchoring at the top and bottom surfaces, the orientation of heliknoton $i$ is coupled to its vertical position $z_i$ by the helical constraint: $\boldsymbol{X}_{i} = (\cos{q z_i},\sin{q z_i},0)$, $\boldsymbol{Y}_{i} = (-\sin{q z_i},\cos{q z_i},0)$ and $\boldsymbol{Z}_{i} = (0,0,1)$, $q = 2\pi/p$. Following \cite{Campos-Villalobos2024}, we express the interaction potential as
\begin{equation}
    \Phi_{IJ} = \sum_{k}^{N_s} w_k G_k,
    \label{eq:lrexpansion}
\end{equation}
where $G_k \equiv G_k(r_{IJ},\Omega_I,\Omega_J,\Omega_{IJ})$ are symmetry functions characterizing the local environment of the interacting particles, $w_k$ are the expansion coefficients, and $N_s$ is the number of terms, chosen to achieve the desired accuracy. The symmetry functions are factorized as
\begin{equation}
G_k = \Lambda_k(r_{IJ})\Psi_k(\Omega_I,\Omega_J,\Omega_{IJ})f_c(r_{IJ}),
\end{equation}
where $\Lambda(r_{IJ})$ and $\Psi(\Omega_I,\Omega_J,\Omega_{IJ})$ encode the radial and orientational dependencies, respectively, and $f_c(r_{IJ})$ is a smooth cutoff function. Following Refs.~\cite{Campos-Villalobos2024,Blum1972,Stone1978}, we write 
\begin{equation}
\Psi_k(\Omega_I,\Omega_J,\Omega_{IJ}) = \prod_j S_{j}(\Omega_I,\Omega_J,\Omega_{IJ})^{a_{kj}},
\end{equation}
where $a_{kj} \in \{0, 1, 2, 3, \cdots \}$ is an exponent and $S_{j}$ are from a set of orthogonal functions known as $S$-functions. For \textit{achiral} interactions with spatial inversion symmetry, only the first $15$ lowest-rank \textit{even} $S$-functions are required. These correspond to the cosines of the angles between the orientation vectors $\{\boldsymbol{X}_I, \boldsymbol{Y}_I, \boldsymbol{Z}_I\}$, $\{\boldsymbol{X}_J, \boldsymbol{Y}_J, \boldsymbol{Z}_J\}$, and $\hat{\boldsymbol{r}}_{IJ} = \boldsymbol{r}_{IJ}/r_{IJ}$. Chiral interactions require another $9$ lowest-rank \textit{odd} S-functions that involve cross-products of these vectors. We define the radial part $\Lambda_k(r_{IJ}) = \exp[-\eta_k(r_{IJ}-R_k)^2]$ and the cutoff $f_c(r_{IJ}) = \left[\cos(\pi r_{IJ}/r_c) + 1\right]/2$ for $r_{IJ} \leq r_c$, and zero otherwise \cite{Behler2007}. Here $\eta_k$ and $R_k$ are tunable parameters, and $r_c$ is the cutoff distance for the interactions.

The construction of the coarse-grained potentials begins with generating a large but manageable pool of candidate symmetry functions by varying $\eta$, $R$, and $a_{j}$ over a suitable range of values. Specifically, $\eta$ was chosen from the set $\{34.6, 6.56, 2.48, 1.24, 0.73, 0.47\}$. For a given $\eta$, the set of $R$ values is given by $R = \left[1.1 + \frac{0.1666}{\sqrt{\eta}}i\right]p,$ where $i \in \{0,1,2,3,\dots\}$, with the constraint $R < r_c$ ($r_c = 6p$). The angular basis functions $\Psi$ were constructed as products of the 15 lowest-rank \textit{even} $S$-functions, raised to positive integer powers $a_j$, with the constraint $\sum_j a_j \leq a_{\max}$. The CG potential is constructed by iteratively adding the symmetry function with the largest correlation to Eq.~(\ref{eq:lrexpansion}) until the desired accuracy is achieved. Full details of the implementation are provided in Refs.~\cite{campos2021machine,Campos-Villalobos2024}. {\color{black}The FG data was split into training and test sets with a typical ratio of 80:20, respectively.} For most cases studied here, setting $a_{\max} = 8$ yielded accurate coarse-grained representations of heliknoton pair interactions, using $N_s = 32$ terms selected from the candidate pool, with root-mean-square errors below $10 k_{\mathrm{B}}T$. For LC-2 at $U = 4.5$V, however, convergence required up to 64 terms with $a_{\max} = 16$.

\begin{figure}
    \centering
    \includegraphics[width=\linewidth]{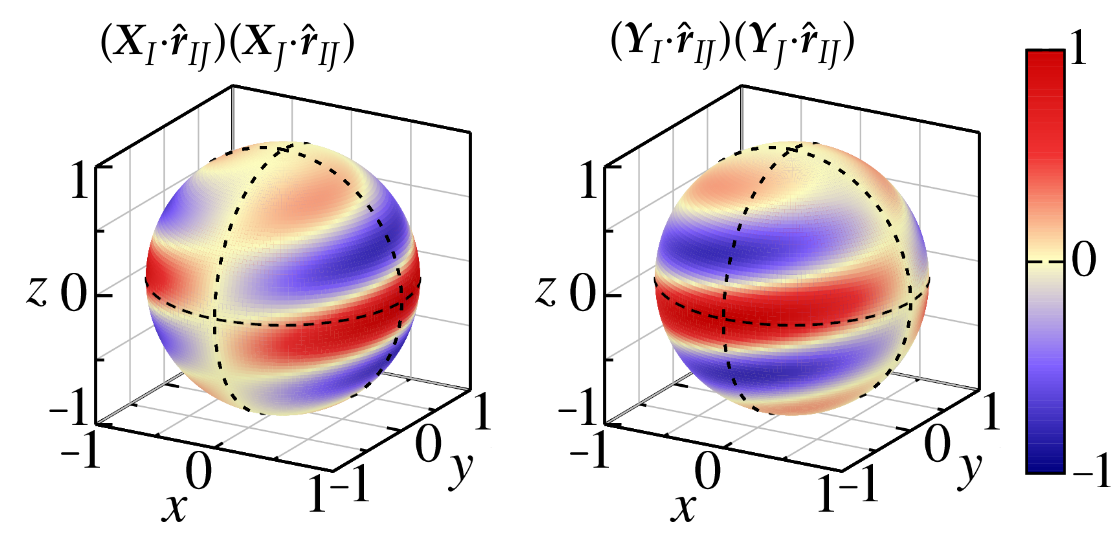}
    \caption{The $S$-function pairs that contribute to the chirality of the CG model, visualized as color maps on the unit sphere defined by $\hat{\boldsymbol{r}}_{IJ}$.}
    \label{figs1}
\end{figure}

The effective number of candidate functions is substantially reduced by the fact that several of the lowest-rank $S$-functions vanish or become constants due to the alignment of heliknotons with the helical background. Interestingly, \textit{odd} $S$-functions were not required, despite the intrinsically chiral nature of the heliknoton textures. In this regard, although the learned heliknoton potentials are constructed from (achiral) dot-product features of the in-plane orientation vectors, such as $(\boldsymbol{X}_{I} \cdot \hat{\boldsymbol{r}}_{IJ})(\boldsymbol{X}_{J} \cdot \hat{\boldsymbol{r}}_{IJ})$ and $(\boldsymbol{Y}_{I} \cdot \hat{\boldsymbol{r}}_{IJ})(\boldsymbol{Y}_{J} \cdot \hat{\boldsymbol{r}}_{IJ})$, the coupling of the heliknoton orientations $\boldsymbol{X}_i$ and $\boldsymbol{Y}_i$ ($i \in \{I,J\}$) with the helical background renders these features chiral {\color{black}[see Fig.~\ref{figs1}]}. Specifically, under the constraints $\boldsymbol{X}_{i} = (\cos{q z_i},\sin{q z_i},0)$, $\boldsymbol{Y}_{i} = (-\sin{q z_i},\cos{q z_i},0)$, where $z_i$ is the vertical position of heliknoton $i$ and $q = 2\pi/p$, it can be shown that the angular basis functions mentioned above gain an odd contribution $\propto \sin(q \Delta z)$, which changes sign when heliknoton $J$ is positioned above ($+\Delta z$) versus below ($-\Delta z$) heliknoton $I$. It also changes sign under a reversal of the handedness ($q \to -q$) of the helical background. Thus, handed interactions emerge naturally and can be fully captured even without explicitly chiral descriptors.

\subsection{Self-Assembly and Dynamical Simulations}
{The self-assembly simulations were performed using the virtual move Monte Carlo algorithm which identifies clusters of interacting particles and moves them collectively based on the gradients of the potential energy~\cite{VMMC,Whitelam2007,Whitelam2009}. As the interaction potentials present sizable energy barriers ($\sim 100k_{\mathrm{B}}T$) in some cases, we employ an annealing protocol in the simulations, starting from a high temperature ($k_{\mathrm{B}}T \sim 50$) and cooling the system slowly to obtain large, defect-free crystallites. Simulations were typically performed for $2 \times 10^6$ MC sweeps (corresponding to $N$ trial moves), extending to $10^7$ MC sweeps in some cases. The 3D heliknoton assembly shown in Fig.~\ref{fig5} was obtained by placing self-assembled monolayers at vertical separations in the range $1.5p$ to $2.5p$, followed by relaxation to identify stable structures.}

{\color{black}The dynamical simulations shown in Fig.~\ref{fig3}B were performed by numerically integrating the overdamped limit of the Langevin equation,
\begin{equation}
\boldsymbol{\dot{r}}_I = -\frac{D}{k_{\mathrm{B}}T}\boldsymbol{\nabla} \Phi_I + \sqrt{2D}~\boldsymbol{\eta}_I(t),
\end{equation}
where $\Phi_I = \sum_J \Phi_{IJ}$, $D$ is the diffusion coefficient (treated as an adjustable parameter to match experimentally observed dynamics) and $\boldsymbol{\eta}_I$ is a Gaussian thermal noise satisfying $\langle\eta_{i\alpha}(t)\eta_{j\beta}(t')\rangle = \delta_{ij}\delta_{\alpha\beta}\delta(t-t')$. We do not consider rotational motion as the heliknoton orientations are fixed in this case, being confined to the vertical mid-plane of the LC cell.}
\clearpage % Clear all remaining figures and tables then start a new page

%

%%%%%%%%%%%%%%%% ACKNOWLEDGEMENTS %%%%%%%%%%%%%%%

\end{document}